\documentclass[twocolumn,aps,prb,floatfix]{revtex4}

\usepackage{graphicx}
\usepackage{amsmath}
\usepackage{latexsym}
\usepackage{amsmath}

\newcommand{\beq}{\begin{equation}}
\newcommand{\eeq}{\end{equation}}

\begin{document}

\title{Scaling properties of the Anderson model in the Kondo regime studied by $\sigma G \sigma W$ formalism}

\author{Catalin D. Spataru}
\email{cdspata@sandia.gov}
\affiliation{Sandia National Laboratories, Livermore, CA 94551, USA}

\begin{abstract}
The symmetric Anderson model for a single impurity coupled to two leads is studied at strong interaction using the GW approximation within the $\sigma G \sigma W$ formalism. We find that the low energy properties
show universal scaling behavior in the asymptotic regime. While the GW scaling functions are similar in form to the scaling functions known from the numerically exact solution, they are characterized by a different parameter value indicating that GW fails to describe correctly spin correlations between the impurity and lead electrons.
We also compare the GW and exact Kondo scales for a broad range of the interaction strength. In contrast to the exponential behavior shown by the exact solution, the GW Kondo scale depends algebraically on the interaction strength.
\end{abstract}

\maketitle

\section{Introduction}
As the size of modern electronic components is pushed toward the molecular limit, the effects of electron-electron interactions can be dramatically enhanced due to quantum confinement. Therefore, understanding strongly correlated phenomena in nanoscale electron transport is important for the development of future electronic devices. 
In this regard, the Anderson impurity model \cite{Anderson} has been very useful, being able to capture the esential physics of complex transport phenomena. Examples include magnetic nanojunctions composed of atoms, molecules or quantum dots \cite{Goldhaber,Cronenwett,McEuen,Park,Glazman} that show a conductance minimum in the linear-response at low temperature, similar in origin to the well-known resistance minimum shown by metals with magnetic impurities: as temperature is lowered, the conductance decreases due to suppressed charge fluctuations (the Coulomb blockade effect), and then increases due to enhanced spin-flip scattering  (the Kondo effect). 

Presently these strongly correlated phenomena cannot be described entirely from first principles. Recent theoretical approaches \cite{Lucignano,Tiago} rely on {\it ab initio}, perturbative methods to extract parameters for a specifically designed Anderson model that is then solved by non-perturbative means such as the numerical renormalization group (NRG) method. Using entirely {\it ab initio} approaches to gain fundamental understanding of strongly correlated electron trasport remains an open challenge even for idealized nanojunctions.

The spin-1/2 single-impurity Anderson model (SIAM) is a basic model in condensed-matter physics and a paradigm for a system of strongly interacting electrons. At equlibrium, its thermodynamic properties can be obtained exactly by means of the Bethe-Ansatz, while its dynamic (excited state) properties can be obtained  numerically with desired precision, using a variety of non-perturbative methods such as NRG and quantum Monte Carlo (QMC). At non-equilibrium, a major step toward solving the symmetric SIAM has been very recently achieved via diagrammatic QMC \cite{Werner}. 

Considering the SIAM as an idealized nanojunction, it is tempting to study its properties in the strong interaction regime using a many-body perturbative approach that is amenable to {\it ab initio} calculations. Progress along this direction could open the door for complete first-principles studies of strongly correlated phenomena in realistic nanojunctions. That such progress is possible is indicated by an important property of the SIAM, namely that the system is always in the Fermi-liquid state as it does not undergo any phase transition; thus, perturbation expansion about the impurity Coulomb interaction $U$ is guaranteed to converge \cite{Horvatic,Anderson_book}. 
 
Treating the dynamic properties of the SIAM with low-order (up to fourth \cite{Yamada,Ueda,Hamasaki}) perturbation expansion about $U$ has provided so far only a qualitative description
of the interesting features that the impurity density of states develops with increasing $U$
(such as the emergence of the sharp Kondo peak near the Fermi level $E_F$ or the formation of Hubbard bands away from $E_F$ near $\pm U/2$), and then only for not too large interaction strength. In particular, low-order expansions in $U$ yield a non-universal \cite{Silver} Kondo resonance which narrows algebraically \cite{Hewson} in $U$ instead of exponentially.

An alternative approach that we consider involves perturbation expansion about the screened Coulomb interaction $W$ \cite{Hedin_W}.   
The first order perturbation expansion about $W$, known as the GW approximation \cite{Hedin} (GWA), is one of the best available tools in describing the quasiparticle properties of real materials \cite{Hybertsen,Godby,Aryasetiawan,Aulbur,Faleev} characterized by not too strong electron correlation. The GWA has been applied also to the SIAM \cite{Thygesen1,Wang,Thygesen2,Spataru}, showing clearly that it fails to reproduce the Hubbard bands unless it artificially breaks spin symmetry. 
However, an important question remains \cite{Wang,Thygesen2}:
how well does GW describe the Kondo peak? 
That this issue is still open is due in part to the fact that a non-magnetic GW solution has been elusive in the strong interaction regime near half-filling and low temperature and bias voltage \cite{Spataru}. Our work answers the above question by calculating such GW solution in the electron-hole symmetric case, 
where charge fluctuations are most suppressed. 

An important conclusion of our work is that GW shows universal scaling behavior in the Kondo regime. A careful analysis of this behavior shows that GW neither describes satisfactorily the Kondo resonance nor the thermodynamic and transport properties that depend on it. On a more positive note, we have identified, employing the recently developed $\sigma G \sigma W$ formalism \cite{Biermann}, a new GW flavor which shows marked improvement over previously used GW flavors; this finding might be useful in applications of GW to moderately correlated electron phenomena.

The rest of the paper is organized as follows. The model Hamiltonian and the underlying $\sigma G\sigma W$ formalism are described in Sec. II. Section III presents results of GW calculations for the in- and out-of-equilibrium low-energy properties of the SIAM in the strong interaction regime, compared  (at equilibrium) to results known from the numerically exact or Bethe-Ansatz solution.
Simple arguments attempting to relate the parameters that control the GW scaling functions and the Kondo scale are provided in the Appendix.

\section{Formalism}

\subsection{Hamiltonian}
We consider the Anderson model \cite{Anderson} for an impurity coupled symmetrically to (left and right) non-interacting semi-infinite leads. 
The non-interacting part of the Hamiltonian describing the spin-1/2 impurity has the form:
\beq
H_{\text{imp}}=\sum_{\sigma} \epsilon_{d} d^\dagger_\sigma d_\sigma
\eeq 
where $d^\dagger_\sigma$ creates an electron with spin $\sigma$ on the impurity. 

We describe the leads within the infinite band-width limit (this approximation has negligible effect on low-energy quantities that are the subject of our paper). The hybridization function describing the coupling between leads and impurity is then just a constant, the effective coupling strength that we denote by $\Delta$. 

The electron-electron interacting part of the Hamiltonian is localized on the impurity and represented by the usual $U$ term:
\beq
H_{\text{e-e}}=U n_{d\uparrow} n_{d\downarrow}=\frac{1}{2}
\sum_{\alpha,\alpha',\beta,\beta'} d^{\dagger}_{\alpha}
d^{\dagger}_{\beta} V_{\alpha  \alpha',\beta \beta'}
d_{\beta'} d_{\alpha'}
\label{H_e_e}
\eeq
where $n_{d\sigma}$ is the electron occupation number of spin $\sigma$ on the impurity and $U$ is the 
repulsive Coulomb interaction between spin up and spin down impurity electrons. 

There are several physically appealing choices for the two-particle electron-electron interaction $V$ that give the same $H_{\text{e-e}}$ in Eq. \eqref{H_e_e}. In this work, we consider two different spin-dependent (and free of self-interaction effects) versions: 
\beq
V^1_{\alpha  \alpha',\beta \beta'}=
- U (1-\delta_{\alpha\beta}) (1-\delta_{\alpha\alpha'}) (1-\delta_{\beta\beta'}) 
\eeq
and: 
\begin{eqnarray}
V^2_{\alpha  \alpha',\beta \beta'}
=
U (1-\delta_{\alpha\beta}) \delta_{\alpha\alpha'} \delta_{\beta\beta'} 
\end{eqnarray}

\begin{figure}
\resizebox{10.0cm}{!}{\includegraphics[angle=180]{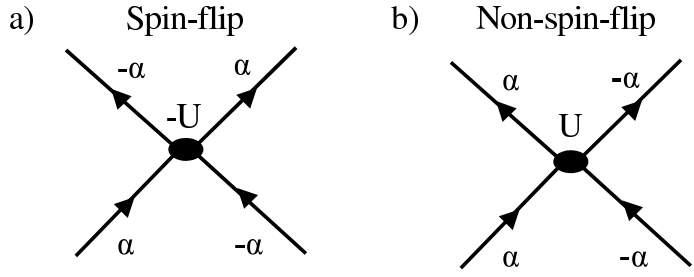}}
\caption{Diagrammatic representation in spin space of the local two-particle Coulomb interaction $V$: (a) spin-flip case, $V^1_{\alpha  \alpha',\beta \beta'} = - U (1-\delta_{\alpha\beta}) (1-\delta_{\alpha\alpha'}) (1-\delta_{\beta\beta'})$, and (b) non-spin-flip case, $V^2_{\alpha  \alpha',\beta \beta'} = U (1-\delta_{\alpha\beta}) \delta_{\alpha\alpha'} \delta_{\beta\beta'}$ .}
\label{V_diagram}
\end{figure}

We use $V^1$ and $V^2$ in the context of the GWA: the newly applied $V^1$ describes spin-flip scattering of an electron upon interaction with an opposite spin electron, as shown schematically in Fig. \ref{V_diagram}(a); $V^2$ does not allow such spin-flip scattering [see Fig. \ref{V_diagram}(b)] but has been found superior \cite{Thygesen1,Wang,Thygesen2,Spataru,Verdozzi} over the spin-independent  version $V^3_{\alpha  \alpha',\beta \beta'}
= U \delta_{\alpha\alpha'} \delta_{\beta\beta'}$ ($V^3$ is the form most commonly used in {\it ab initio} GW calculations of real materials). In the next section we show that $V^1$ outperforms $V^2$ when describing the low-energy properties of the SIAM.

\subsection{
$\sigma G \sigma W$ formalism}
 
We consider the SIAM in both equilibrium and nonequilibrium situations using a nonequilibrium Green's functions (NEGF) approach. Electron correlation effects are obtained by solving self-consistently for the various (retarded, lesser, etc ..) Green's functions and self-energies of the impurity. Details of our NEGF approach can be found in Ref. \cite{Spataru};
for simplicity we show in this section only time-ordered quantities relevant for the equilibrium, zero temperature case.

At half-filling (one electron on the impurity) the impurity orbital energy is  $\epsilon_{d}=-\frac{U}{2}$ (the zero of energy is chosen such that the chemical potential is $\mu=0$). Then, the impurity Green's function reads:
\beq
G_\sigma(\omega)=\frac{1}{\omega-\left[-U/2+V^H_\sigma+\Sigma_\sigma(\omega)\right]+i \Delta \text{ sgn}(\omega)}
\eeq
Here, $V^H_{\sigma}$ is the spin-dependent Hartree potential \cite{Biermann}; one has $V^H_\sigma=0$ in the spin-flip case and $V^H_\sigma=U \langle n_{-\sigma}\rangle$ in the non-spin-flip case.

The impurity self-energy $\Sigma$ is evaluated using the self-consistent GWA within the $\sigma G \sigma W$ formalism \cite{Biermann}, which is based on the recent generalization of Hedin's equations for quantum many-body systems with spin-dependent interactions. 
Following Ref. \cite{Biermann}, we decompose $V$ in a basis formed by the Pauli matrices: 
\beq
V_{\alpha  \alpha',\beta \beta'}=\sigma^I_{\alpha\alpha'} V_{IJ} \sigma^I_{\beta\beta'}
\eeq
where repeated indices are summed and subscripts denoted by capital letters (e.g. $I,J$) are indices representing the four Pauli matrices $\sigma^0$,$\sigma^x$,$\sigma^y$ and $\sigma^z$.
The $V_{IJ}$ components associated with $V^1$ and $V^2$ are easily obtained from:
\beq
V^1_{\alpha  \alpha',\beta \beta'}=
-\frac{U}{2}\left(\sigma^x_{\alpha\alpha'} \sigma^x_{\beta\beta'}
+\sigma^y_{\alpha\alpha'} \sigma^y_{\beta\beta'}\right)
\label{V_spin_flip}
\eeq
and
\beq
V^2_{\alpha  \alpha',\beta \beta'}
=\frac{U}{2}\left(\sigma^0_{\alpha\alpha'} \sigma^0_{\beta\beta'}
-\sigma^z_{\alpha\alpha'} \sigma^z_{\beta\beta'}\right)
\label{V_non_spin_flip}
\eeq
which shows that both $V^1$ and $V^2$ have contribution from the 
spin-channel. (In contrast, the spin-independent version $V^3$ has contribution only from the 
charge channel: $V^3_{00}=U$, $V^3_{I\neq0 J\neq0}=0$.)

Within the $\sigma G \sigma W$ formalism the spin-dependent Hedin's equations read \cite{Biermann}
\beq
\Sigma_\sigma(\omega)= i \int \frac{dE}{2\pi} e^{i0^+E} ~\sigma^I_{\sigma \gamma} G_\gamma(E) \sigma^J_{\gamma \sigma} W_{JI}(\omega-E)
\nonumber
\eeq
\beq
W_{JI}(\omega)=V_{IJ}+V_{IK}P_{KL}(\omega)W_{LJ}(\omega)
\label{genHedin_SW}
\eeq
where the irreducible polarization is evaluated within random-phase approximation (RPA),
\beq
P_{IJ}(\omega)= -i \int \frac{dE}{2\pi} \sigma^I_{\alpha \beta} G_\beta(E) \sigma^J_{\beta \alpha}G_\alpha(E-\omega)
\label{genHedin_P}
\eeq

The spin-flip and non-spin-flip GW flavors are obtained using definitions \eqref{V_spin_flip} and \eqref{V_non_spin_flip} for $V_{IJ}$ in Hedin's Eq. \eqref{genHedin_SW}. Using the following notation for the polarization bubble in spin space:
\beq 
\tilde{P}_{\alpha,\beta}(\omega)\equiv-i \int \frac{dE}{2\pi} G_\alpha(E) G_\beta(E-\omega)
\eeq
the final expression for $\Sigma$ within the spin-flip GW flavor reads
\beq
\Sigma^1_{\sigma}(\omega)= -i \int \frac{dE}{2\pi} e^{i0^+E} ~G_{-\sigma}(E) 
\nonumber
\eeq
\beq
\times 
\frac{U+U^2\tilde{P}_{-\sigma, \sigma}(\omega-E)}
{1+U\tilde{P}_{-\sigma,\sigma}+U\tilde{P}_{\sigma,-\sigma}+U^2\tilde{P}_{-\sigma,\sigma}\tilde{P}_{\sigma,-\sigma}}
\eeq
(above, we omitted for simplicity the energy variable on $\tilde{P}$ showing in the denominator).

In the spin-flip case the expression for the self-energy of an electron with spin $\sigma$ involves a Green's function for an electron with opposite spin $-\sigma$, and similarly the polarization bubble involves Green's functions of electrons with opposite spin. 
The same does not hold in the non-spin-flip case, where the self-energy expression reads \cite{Wang}
\beq
\Sigma^2_{\sigma}(\omega)= i \int \frac{dE}{2\pi} ~e^{i0^+E} ~G_{\sigma}(E) 
\nonumber
\eeq
\beq
\times
\frac{U^2\tilde{P}_{-\sigma,-\sigma}(\omega-E)}
{1-U^2\tilde{P}_{\sigma,\sigma}(\omega-E)\tilde{P}_{-\sigma,-\sigma}(\omega-E)}
\eeq

In the absence of a magnetic field, the exact solutions for $G$ and $\Sigma$ are non-magnetic, independent of spin. Similarly, we consider in this work only non-magnetic GW solutions in which case the expressions for $\Sigma^1$ and $\Sigma^2$ simplify and read (we omit from now on spin indices):
\beq
\Sigma^1(\omega)= -i \int \frac{dE}{2\pi} ~e^{i0^+E} ~G(E) 
\frac{U}
{1+U\tilde{P}(\omega-E)}
\eeq

\beq
\Sigma^2(\omega)= i \int \frac{dE}{2\pi} ~e^{i0^+E} ~G(E) 
\frac{U^2\tilde{P}(\omega-E)}
{1-U^2\tilde{P}^2(\omega-E)}
\eeq
 
\section{Results}

A well known characteristic of the SIAM is that its low-energy physics shows universal scaling behavior in the Kondo regime, i.e. near half-filling  and for effective interaction strength:
\beq
u\equiv\frac{U}{\pi\Delta}>2
\eeq
(a strong-interaction asymptotic regime is approached exponentially fast for $u>2$). A signature of this aspect is that functions describing spectroscopic, thermodynamic, and transport quantities at low energies become independent of $u$ when their energy argument is scaled by the Kondo temperature $T_K$. 

\subsection{Scaling behavior of the inpurity density of states}
\begin{figure}
\resizebox{8.0cm}{!}{\includegraphics{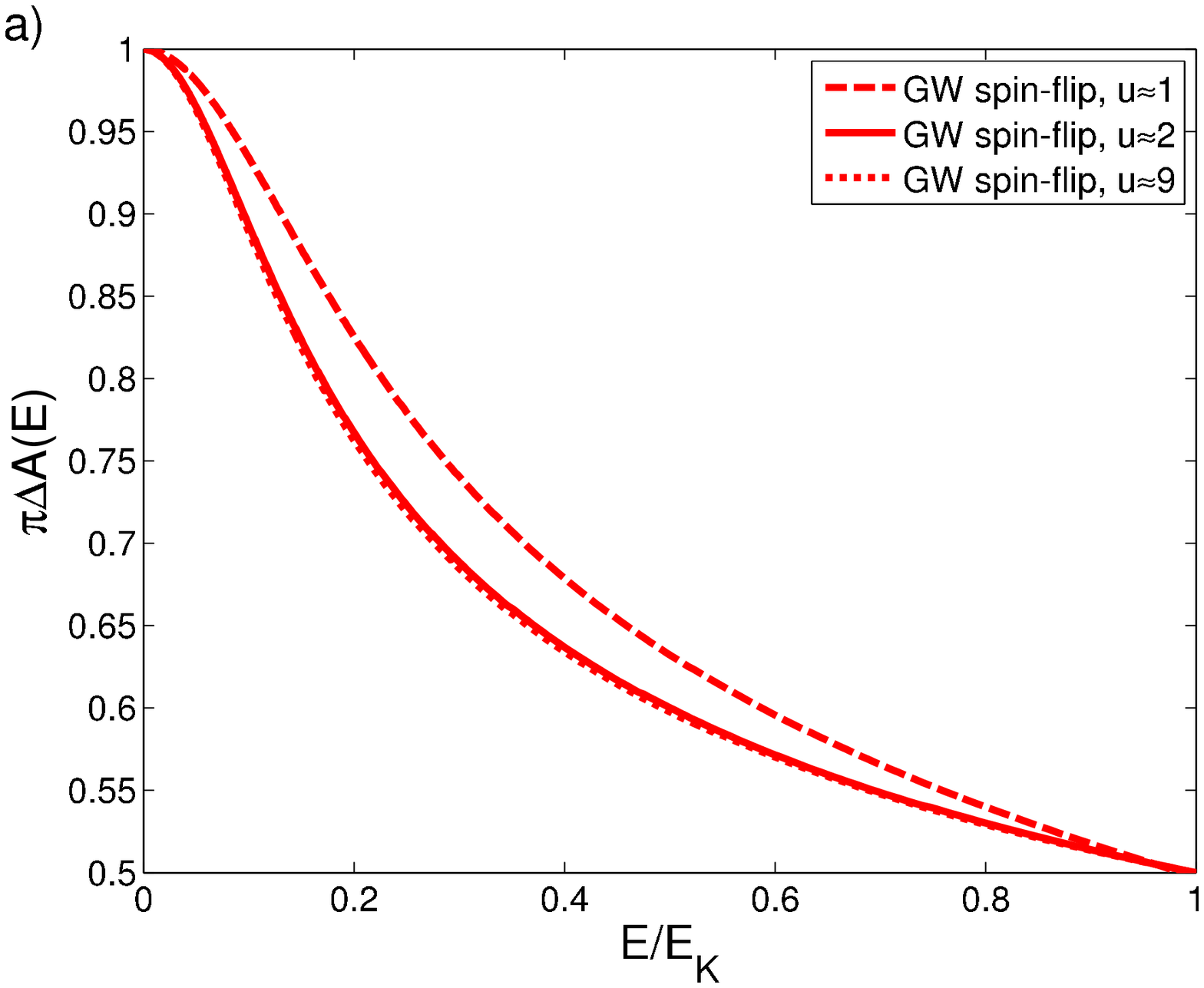}}
\resizebox{8.0cm}{!}{\includegraphics{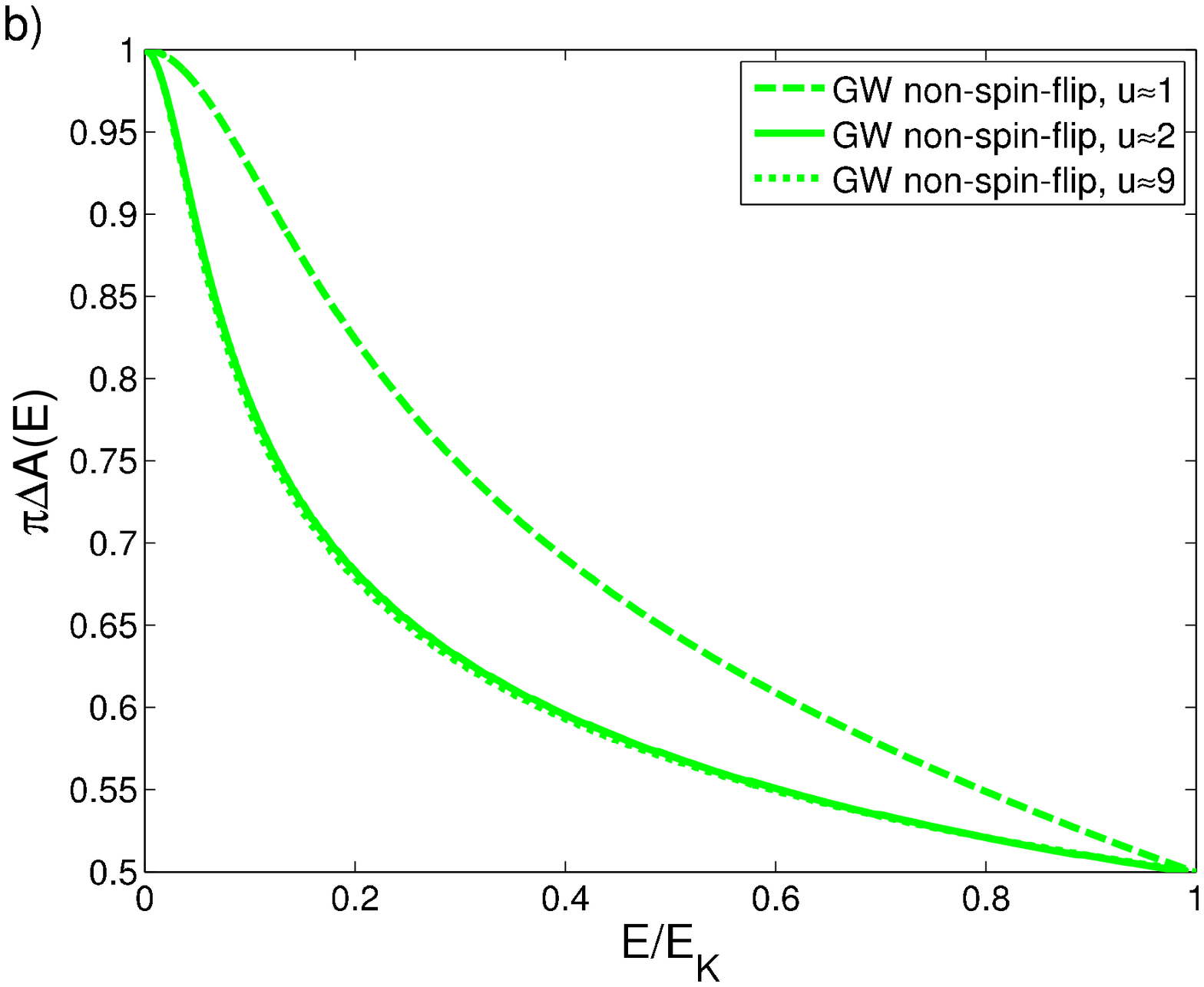}}
\resizebox{8.0cm}{!}{\includegraphics{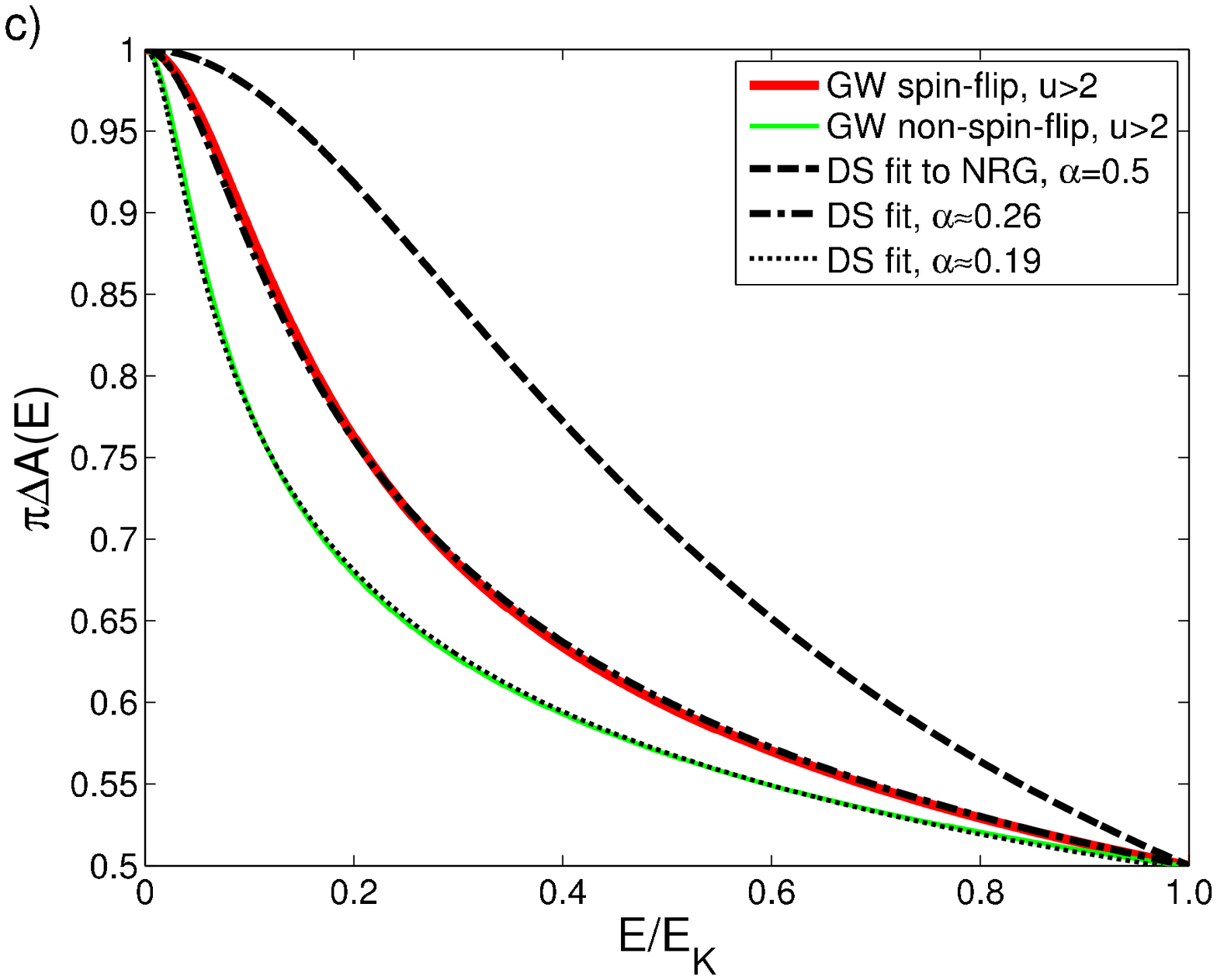}}
\caption{The impurity density of states at low energy and $T=0$, $V=0$.
(a) Results within the spin-flip flavor of spin-dependent GW, for several values of the effective interaction strength $u$. (b) Results within the non-spin-flip flavor of spin-dependent GW, for the same $u$ values as in (a). (c) Comparison between spin-dependent GW and the heuristic Doniach-Sunjic (DS) fit to NRG results in the asymptotic regime.
}
\label{A_univ}
\end{figure}

Figure \ref{A_univ} shows results at $T=0$ for the impurity density of states:
\beq
A(E)\equiv \frac{1}{\pi}\left |\Im G(E)\right |
\eeq
 as function of energy divided by the
half-width 
at half-maximum of the central peak (denoted $E_{\text{K}}$). 
Figure \ref{A_univ}(a) shows results 
obtained 
within the spin-flip GW flavor, while Fig. \ref{A_univ}(b) treats the non-spin-flip 
case. We see that GW captures the aformentioned universality aspect:
for $u>2$, $A(E)$ for various $u$ are almost undistinguishable.

From the numerically exact solution it is known that the low-energy part of
 the impurity density of states can be fitted very well with the following form \cite{Frota,Rosch}:
\beq 
A^{\text{DS}}(E)= \frac{1}{\pi\Delta} \Re\left\{\left(\frac{i\Delta_ {\text{DS}}}{E+i\Delta_ {\text{DS}}}\right)^\alpha\right\}
\label{DS}
\eeq
with $\Delta_ {\text{DS}}/E_{\text{K}}$ a constant determined by $\alpha$ via $A^ {\text{DS}}(E_{\text{K}})=1/(2\pi\Delta)$. 

The heuristic form shown in Eq. \eqref{DS} is inspired by the Doniach-Sunjic law \cite{Doniach} that describes photoemission of core electrons by x-ray in metals. The role of the core electron in metals must be reflected in the SIAM by some sort of spin polarization cloud \cite{Affleck} developed from electrons in the leads screening (at low $T$ and low $V$) the unpaired impurity spin.

For the numerically exact solution of the SIAM the best Doniach-Sunjic fit is realized for:
\beq
\alpha_{\text{exact}}=0.5
\eeq 
This is
believed to arise from a Fermi-edge singularity \cite{Fermi_edge} in the presence of a $\pi/2$ phase-shift at the Fermi level. 
The scaling function associated with $A(E)$ corresponding to the numerically exact solution is shown in Fig. \ref{A_univ}(c) as the black dashed curve.  

Also shown in Fig. \ref{A_univ}(c) are the GW scaling functions representative of $A(E)$ in the asymptotic regime: the thick solid (in red online) curve represents the spin-flip case and the thin solid (in green online) curve the non-spin-flip case. The Doniach-Sunjic fit works very well in the GW case as well (with an error  $\pi\Delta\left|A^ {\text{DS}}(E)-A(E)\right| \ll 1$), as demonstrated in 
Fig. \ref{A_univ}(c) by the black discontinuous curves. More precisely
\begin{eqnarray}
\alpha_{\text{GW}}=
\left\{
  \begin{array}{ l l}
     0.26,       & \text{spin-flip case}  \\
     0.19,       & \text{non-spin-flip case.}  
  \end{array}
\right.
\end{eqnarray}  

We see that spin-flip GW does better than non-spin flip GW in describing the shape of the scaling function associated with $A(E)$. However, the smallness of $\alpha_{\text{GW}}$ relative to $\alpha_{\text{exact}}$ indicates that both spin-dependent GW flavors strongly overestimate spin correlations between impurity and lead electrons.  [In contrast, within spin-independent GW (or, more obviously, the Hartree-Fock approximation) which does not account for correlations in the spin channel, the Doniach-Sunjic fit is trivially realized for $\alpha=1$.]

\subsection{Scaling behavior of the linear-response conductivity}

\begin{figure}
\resizebox{10.0cm}{!}{\includegraphics{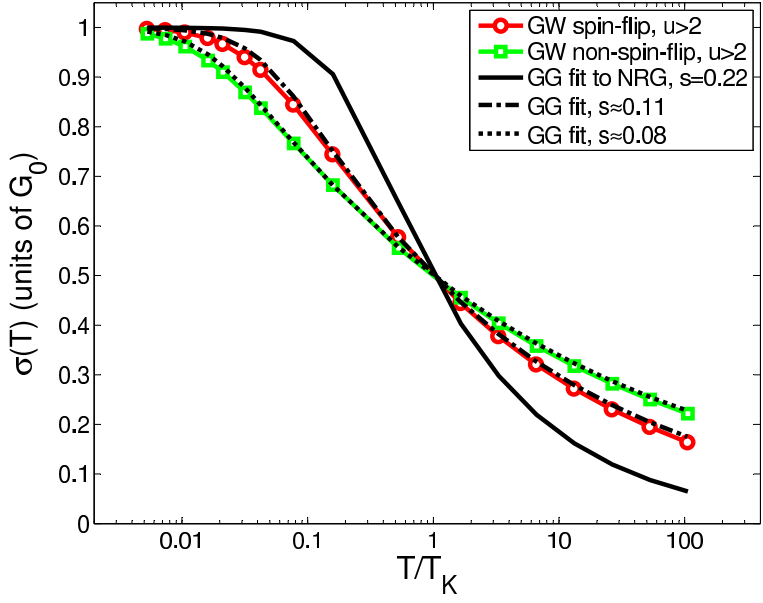}}
\caption{Linear-response conductivity in the asymptotic regime at low temperature, calculated within spin-dependent GW and from the empirical Goldhaber-Gordon (GG) fit to NRG results.}
\label{G_T_univ}
\end{figure}

Figure \ref{G_T_univ} shows the linear-response conductivity
\beq
\sigma(T)\equiv  \left.\frac{\partial I}{\partial V} \right\vert_{V=0}
\eeq
in units of the quantum conductance $G_0\equiv 2e^2/h$ and evaluated as:
\beq
\frac{\sigma(T)}{G_0}=\pi\Delta\int_{-\infty}^\infty dE \frac{-\partial f(E)}{\partial E} A(E)
\eeq
where $f(E)=1/[{\text exp}(E/T)+1]$ is the usual statistical factor for a system of electrons in equilibrium at finite temperature $T$ and $\mu=0$.

Universal curves are obtained in the asymptotic regime $u>2$ by plotting $\sigma(T)$ as a function of $T/T_K$ where $T_K$ is the Kondo temperature.  There are several definitions of $T_K$ existing in the literature, given that the Kondo scale can be defined in the asymptotic regime only up to a multiplicative factor. Here, we define $T_K$ as the half-width at half-maximum of the linear response conductivity:  $\sigma(T_K)=0.5 G_0$. 

The numerically exact, universal results for $\sigma(T)$ are represented in Fig. \ref{G_T_univ} by the black solid curve, obtained here from the empirical Goldhaber-Gordon form \cite{GG}:
\beq 
\frac{\sigma^{\text{GG}}(T)}{G_0}=
\left( 
\frac{T_K^{'^2}}{T^2+T_K^{'^2}} 
\right)^s
\eeq
where $T_K^{'}=T_K/\sqrt{2^{1/s}-1}$ and the parameter $s$ takes the value $s_{\text{exact}}=0.22$ when the Goldhaber-Gordon form is used to fit NRG results \cite{Costi}.

GW also displays universal scaling behavior for $\sigma(T)$. The curves in red and green  in Fig. \ref{G_T_univ} are representative of the spin-flip and non-spin-flip flavors respectively.
We find that the Goldhaber-Gordon fit works for GW as well, as demonstrated by the black discontinuous curves in Fig. \ref{G_T_univ}. We also find that the value of the parameter $s$ ($s_{\text{GW}}\approx0.11$ in the spin-flip case, and $s_{\text{GW}}\approx0.08$ in the non-spin-flip case) is simply related to the $\alpha$ parameter through:
\beq
\frac{s_{\text{GW}}}{s_{\text{exact}}}=\frac{\alpha_{\text{GW}}}{\alpha_{\text{exact}}}
\eeq

For $T\ll T_K$, all curves in Fig. \ref{G_T_univ} show Fermi-liquid 
behavior, $1-\sigma(T)/G_0\sim (T/T_K)^2$. Due to the smallness of $s$ (or equivalently of $\alpha$), $\sigma$ drops too fast with $T/T_K$ within GW, even though  the spin-flip flavor improves significantly over the non-spin-flip one. For $T\gg T_K$, $\sigma$ drops too slowly, as expected since GW cannot capture the Coulomb blockade regime unless it artificially breaks spin symmetry.

\subsection{Scaling behavior of the differential conductivity}
\begin{figure}
\resizebox{10.0cm}{!}{\includegraphics{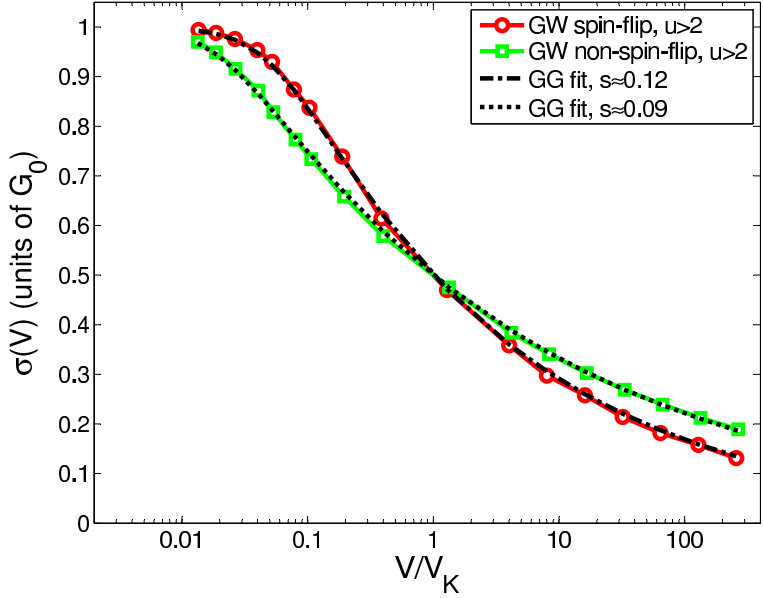}}
\caption{Differential conductivity in the asymptotic regime at low bias voltage and $T=0$, calculated within spin-dependent GW.}
\label{G_V_univ}
\end{figure}

Figure \ref{G_V_univ} shows  GW results for the differential conductivity at $T=0$:
\beq
\sigma(V)\equiv  \left.\frac{\partial I}{\partial V} \right\vert_{T=0}
\eeq
calculated from:
\beq
\frac{\sigma(V)}{G_0}=\pi\Delta A\left(\frac{V}{2}\right)+\pi\Delta\int_{-V/2}^{V/2} dE \frac{\partial A(E)}{\partial V}
\label{sigmaV}
\eeq 
and plotted for $u>2$ as a function of a symmetrically-applied bias voltage $V$ divided by a Kondo scale $V_K$ defined as $\sigma(V_K)=0.5 G_0$. 

The GW results for $\sigma(V)$ show again universal scaling behavior,  i.e. there is just one representative curve for any $u>2$. The curves in red (with circles) and green (with squares) shown in Fig. \ref{G_V_univ} correspond to 
the spin-flip and  non-spin-flip flavors respectively. (We are not aware of numerically exact results that can be compared to our GW data in Fig. \ref{G_V_univ}.)

We find that the GW calculated $\sigma(V)$ can be fitted very well by the Goldhaber-Gordon form $\sigma^{\text{GG}}(V)$, as demonstrated by the discontinuous black curves in Fig. \ref{G_V_univ}. 
The value of $s$ is $\approx 5\%$ larger than the corresponding value in the case of $\sigma(T)$, indicating that
 the effect of a low bias voltage is quite similar (but not exactly the same \cite{terms_sigmaV}) to that of an effective 
temperature.

\subsection{Kondo scale}
\begin{figure}
\resizebox{10.0cm}{!}{\includegraphics{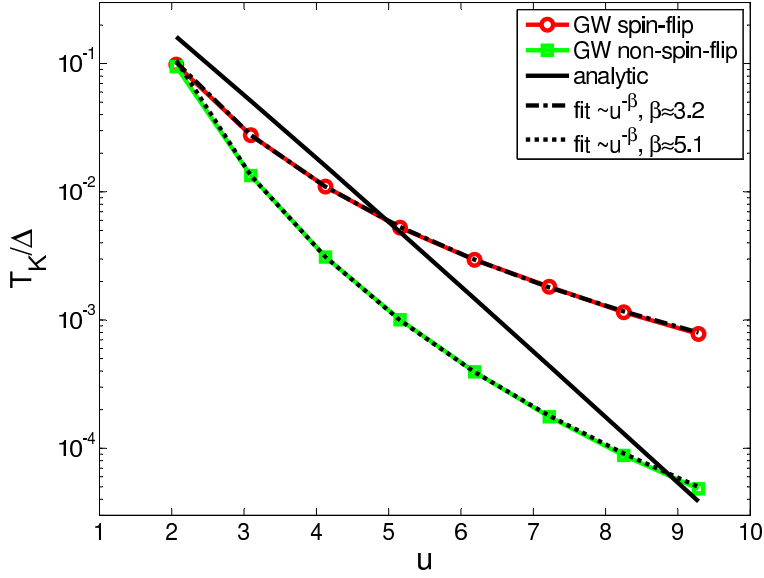}}
\caption{Comparison between the Kondo temperature calculated within spin-dependent GW and the one from the (asymptotically exact) analytic solution (Ref. \cite{prefactor}) for a large range of the effective interaction strength $u$ in the asymptotic regime.}
\label{Kondo_scale}
\end{figure}

We now turn our attention to the Kondo scale that goes inside the scaling functions and compare in Fig. \ref{Kondo_scale} the GW and analytic (asymptotically exact) Kondo temperature for a broad range of the interaction strength in the asymptotic regime.
 
The Kondo temperature of the exact solution is well known \cite{Haldane} to decrease exponentially with $u$,
\beq
\frac{T_K}{\Delta}\sim  u ~e^{{-\pi^2 u/8}}
\label{Kondo_scale_exact}
\eeq
and it is shown as the black solid line in Fig. \ref{Kondo_scale}.
Also shown by the curves in red (with circles) and green (with squares) are the GW results obtained within the spin-flip and non-spin-flip flavor respectively. Importantly, we find that the GW Kondo scale depends algebraically on $u$:
\beq 
\frac{T_K^{}}{\Delta}\sim 
u^{-\beta}
\label{algebraic}
\eeq
 as indicated by the black discontinuous curves in Fig. \ref{Kondo_scale}. 
The value of the parameter $\beta$ that controls the GW Kondo scale is:
\begin{eqnarray}
\beta \approx
\left\{
  \begin{array}{ l l}
     3.2,       & \text{spin-flip case}  \\
     5.1,       & \text{non-spin-flip case}  \hspace{1 cm}
  \end{array}
  \right.
\end{eqnarray}

The large value of $\beta$ implies that the GW Kondo scale decreases fast with increasing interaction strength. For this reason, finding a non-magnetic GW solution for $u>2$ requires a very fine sampling of the energy axis near $E_F$, that we achieve using a discretized logarithmic scale 
$\sim\pm\Lambda^{-n}$
($\Lambda$ as small as $1.005$ and $n$ as large as $4000$ were used for the largest $u$ considered).
In the Appendix we provide some simple arguments that relate $\beta$ to the parameter $\alpha$ that controls the GW scaling functions. 

We also note from Fig. \ref{Kondo_scale} that  spin-flip GW performs better than non-spin-flip GW for the most physically relevant $u$-values (i.e., $u \lesssim 6$).
At strong 
enough interaction both GW flavors are guaranteed to severely overestimate the Kondo 
scale due to their algebraic $u$ dependence.

We conclude this section with the relationship between the Kondo scales $E_\text{K}$, $T_K$ and $V_K$ that go
inside the scaling functions associated with $A(E)$, $\sigma(T)$ and $\sigma(V)$. 
Within both  spin-dependent GW flavors considered we find:
\beq
E_\text{K}^{\text{GW}}\approx 1.2 ~T_K^{\text{GW}}
\eeq
and:
\beq
V_K^{\text{GW}}\approx 1.5 ~T_K^{\text{GW}}
\eeq  
while for the numerically exact solution, NRG yields (see Ref. \cite{Micklitz}): 
\beq
E_\text{K}^{\text{exact}}\approx 2.3 ~T_K^{\text{exact}}.
\eeq
(We are not aware of numerically exact results for $V_K/T_K$.)

\subsection{Fermi-liquid properties and the linear coefficient of the specific heat}

The Fermi-liquid properties manifest themselves at very low energies: $\omega , T,V \ll T_K$. Here, $\Im\Sigma$ is characterized by quadratic behavior, and in the asymptotic regime the impurity self-energy 
has the following low-order expansion
\beq 
\Sigma(\omega,T,V) \approx -\frac{\omega}{Z} - \frac{iC}{2\Delta Z^2}
\left(
\omega^2+\pi^2 T^2+\frac{3}{4}V^2
\right).
\label{FL}
\eeq
For the exact solution, the form shown in Eq. \eqref{FL} can be derived via the Ward identity, and one finds \cite{Yamada,Yosida,Oguri,Kopietz} $C=1$. Within the GWA the Ward identity is not satisfied \cite{Takada} and we find instead $C\approx 2.6$ in the spin-flip case and $C\approx 3.5$ in the non-spin-flip case. 

The Fermi-liquid behavior is dictated by the inverse renormalization factor
\beq
Z^{-1}\equiv 1-\left.\frac{\partial \Sigma}{\partial \omega}\right|_{\omega,T,V=0}
\eeq
which is directly related to the $T$-linear coefficient of the impurity heat capacity
\beq
\gamma\equiv\lim_{T\rightarrow 0}
\left.\frac{1}{T}\frac{\partial E_{\text{imp}}}{\partial T}\right|_{V=0},
\label{gamma_E}
\eeq
where $E_{\text{imp}}$ is the average energy associated with the presence of the impurity \cite{KSS,Spataru}. Indeed, normalizing $\gamma$ to its value in the absence of Coulomb interactions
\beq
\tilde{\gamma}\equiv\frac{3\Delta}{2\pi}\gamma
\eeq
one obtains that $\gamma$ is enhanced by Coulomb interactions by a factor equal to $Z^{-1}$ [we have checked this statement in the GW case by directly evaluating the right-hand side of Eq. \eqref{gamma_E}],
\beq
\tilde{\gamma}=Z^{-1}.
\label{Z_gamma}
\eeq

\begin{figure}
\resizebox{10.0cm}{!}{\includegraphics{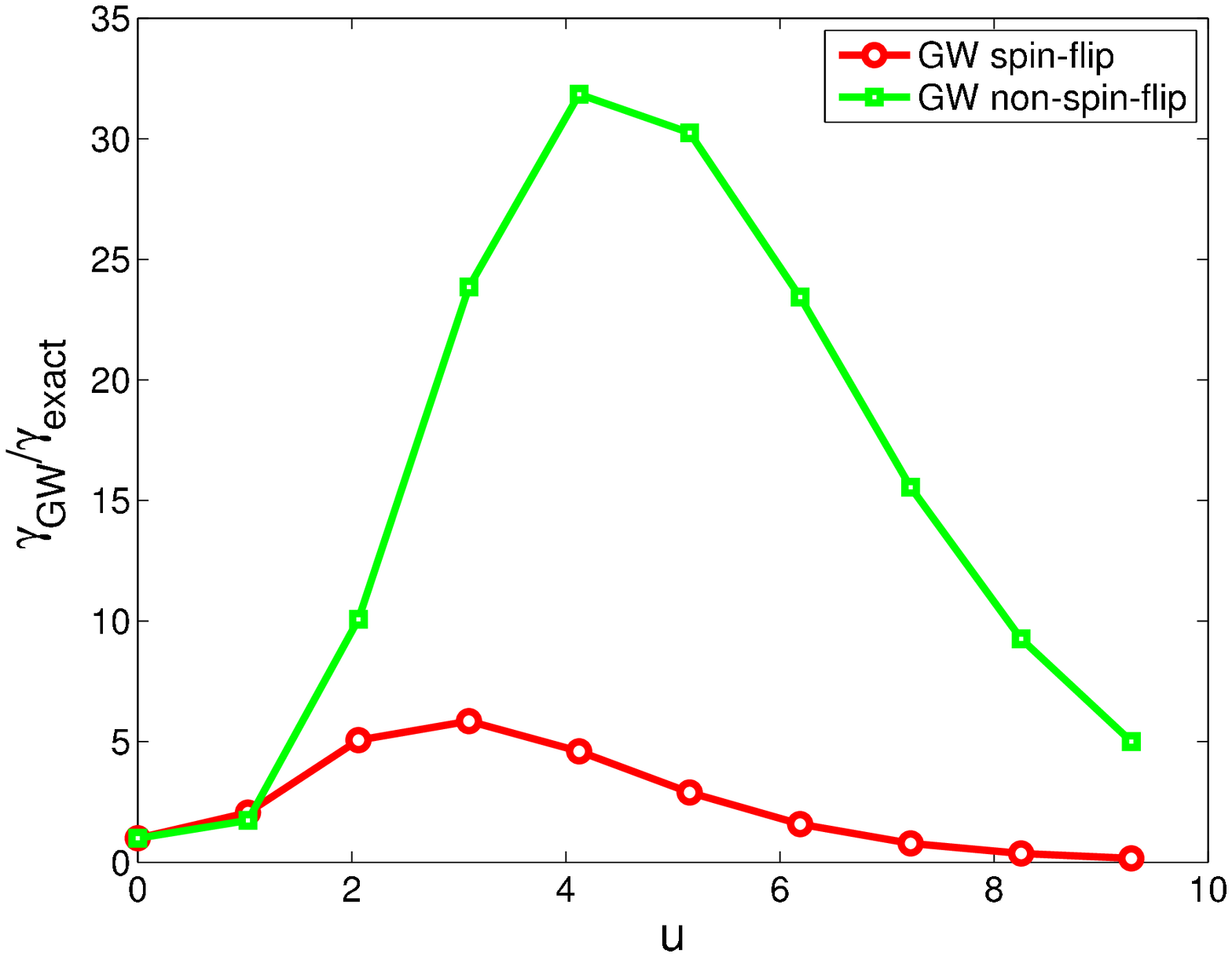}}
\caption{Ratio of the impurity heat capacity calculated within spin-dependent GW to the one calculated exactly (by the Bethe-Ansatz method) for a large range of the effective interaction strength $u$.}
\label{Z_ratio}
\end{figure}

Figure \ref{Z_ratio} shows the ratio ${\gamma}_{\text{GW}}/{\gamma}_\text{exact}$ 
as a function of u, for both GW flavors considered in this work ( ${\gamma}_\text{exact}$ is evaluated according to Ref. \cite{Zlatic}). 
One can see that within non-spin-flip 
GW, ${\gamma}$ is largely overestimated over the entire $u$-range considered; the 
spin-flip GW version significantly improves ${\gamma}$ for $u\lesssim9$, but the 
description remains unsatisfactory. In the limit of very large $u$ both GW flavors are bound to underestimate ${\gamma}$ since universality  implies $Z^{-1} \sim \Delta/T_K$. 

To analyze more in detail the reasons behind this incorrect description, we note that within spin-dependent GW
\beq
Z^{-1}_{GW}\approx \int_{-\infty}^0 dE ~A(E) \frac{d}{dE} \Re W(E) \approx A(0)W(0),
\eeq
where we used the facts that within RPA (i) the energy dependence of $W$ is dictated by a pole situated on the imaginary axis \cite{Hamann}
\beq
W(\omega\rightarrow0) \sim
\frac{\Delta^2}{i\omega-\omega_{SF}} 
\eeq
at the so-called spin-fluctuation frequency $\omega_{SF}$ that is directly related to the static polarization bubble $\tilde{P}(0)$, \cite{White} 
\beq
\omega_{\text{SF}} = \frac{\pi \Delta^2}{U} \left[1+U \tilde{P}(0)\right]
\label{SF_freq}
\eeq
and (ii) we find that $\omega_{\text{SF}}\ll E_{\text{K}}$.

To this end one can write:
\beq
Z^{-1}_{GW} \sim \frac{\Delta}{\omega_{\text{SF}}}
\eeq 
and conclude that the incorrect description of $Z^{-1}_{GW}$ is 
intimately related to the incorrect description of the spin-fluctuation frequency  $\omega_{\text{SF}}$
[or of the static polarization bubble $\tilde{P}(0)$] within RPA. 

The above analysis suggests that vertex corrections beyond RPA in the irreducible polarization $P$ 
are important for a correct description of spin fluctuation effects and of the Kondo peak. 
This complements a finding of previous work \cite{Wang}, namely that vertex corrections beyond GW in the electron self-energy $\Sigma$ are critical for a correct description of charge fluctuation effects and of the  Hubbard bands.

\section{Summary and conclusions}
We have found within GW that the low-energy properties of the SIAM display universal scaling behavior in the Kondo regime. This is remarkable if one thinks of GW as a theory based on perturbation expansion 
about the impurity Coulomb interaction $U$, but perhaps not very surprising given that 
GW corresponds to a partial resummation of an infinite set of such
diagrams. 

We have shown that the scaling functions describing quantities such as the 
impurity density of states or the linear-response conductivity have a similar form 
within GW as in the numerically exact solution, but they are characterized by a different parameter 
value. The deviation of this value from the correct one indicates the extent at which spin-dependent GW overestimates spin 
correlations between the impurity spin and the electrons in the leads.
We have also found that the Kondo scale that goes inside the GW scaling 
functions depends algebraically on the interaction strength, as opposed to 
the exponential dependence shown by the exact solution. 
Both the parameter value 
characterizing the scaling functions and the Kondo scale at physically relevant 
interaction strengths are best described within the 
spin-flip GW flavor. However,  
none of the GW flavors describes satisfactorily the Kondo regime. 

The importance of our analysis should be two-fold. First, it clearly quantifies the extent at 
which various GW flavors fail to describe satisfactorily 
the Kondo regime of the Anderson model.  Second, it suggests that a systematic 
study of the low-energy properties of the Anderson model is possible within a perturbative approach based on expansion about the screened Coulomb interaction $W$: with increasing order of perturbation expansion about $W$, the parameters controlling the scaling functions should converge toward the exact values, while the Kondo scale should improve overall with best accuracy expected at lower $u$. Along this line, extending Hedin's approach up to second order expansion about $W$ appears naturally as a path towards improvement beyond GW/RPA.

\begin{acknowledgments}
I would like to thank Andrew Millis for useful suggestions and Silke Biermann for an instructive discussion of the material in Ref. \cite{Biermann}. Sandia is a multiprogram laboratory
operated by Sandia Corporation, a Lockheed Martin Co., for the United States Department
of Energy under Contract No. DE-AC01-94-AL85000.
\end{acknowledgments}

\section*{Appendix: Simple arguments relating the parameters that control the GW scaling functions and the Kondo scale.}
\renewcommand{\theequation}{A-\arabic{equation}}
\setcounter{equation}{0}

Within GW, we find that the low- and high-energy parts of the impurity (retarded) Green's function have the following forms in the strong interaction regime:
\beq
G(E)\approx
\left\{
  \begin{array}{ l l}
     \frac{1}{i\Delta}
     \left(
     \frac{i\Delta_ {\text{DS}}}{E+i\Delta_ {\text{DS}}}
     \right)^{\alpha}
       & , |E| < E_1   \\ 
     \frac{1}{U} f\left(\frac{E}{U}\right)  & , |E| > E_2  \hspace{0.5 cm} 
     
  \end{array}
  \right.
  \label{low_high}
\eeq
where $E_1\gg T_K$ and $E_2 \ll U$.

The low-energy ($|E|<E_1$) behavior of $G(E)$  is consistent with the Doniach-Sunjic form for the impurity spectral function [see Eq. \eqref{DS}]. With good approximation we find that 
$E_1$ can be pushed up to energies of order $\Delta$.

The high-energy ($|E|>E_2$) behavior of $G(E)$ is consistent with (if not a consequence of) the fact that within RPA one has (to leading order in $1/U$)
\beq
\tilde{P}(0)= -\frac{1}{\pi} \int_{-\infty}^0 dE ~\Im\left\{G^2(E)\right\} \approx -\frac{1}{U},
\eeq
a result intimately related to the smallness of the GW Kondo scale [see Eq. \eqref{SF_freq}]. With relative good approximation, $E_2$ can be pushed down to energies of order $\Delta$. 

Let us assume for the moment that the low- and high-energy behaviors shown in Eq. \eqref{low_high}  are exactly valid in the intermediate-energy regime near $E\approx \Delta\gg \Delta_ {\text{DS}}$. Then, consistency in the $E$-dependence of $G$  at fixed $U$ implies $f(x)\sim x^{-\alpha}$ for $x\ll 1$, while consistency in the $U$-dependence of $G$ at fixed $E$ yields
\beq
\left(\frac{\Delta_ {\text{DS}}}{\Delta}\right)^{\alpha} \sim \frac{1}{u^{1-\alpha}}.
\eeq
Since [see Eqs. \eqref{DS} and \eqref{algebraic}]
\beq
\frac{\Delta_ {\text{DS}}}{\Delta}\sim \frac{T_K}{\Delta} \sim u^{-\beta},
\eeq
one obtains within the above assumption that:
\beq
\beta\approx{1/\alpha}-1.
\eeq 

In reality, the low- and high-energy forms shown in Eq. \eqref{low_high} hold only approximately near $E\approx\Delta$ and the relation between
 the parameter $\beta$ controling the GW Kondo scale and the parameter $\alpha$ controling the GW scaling functions is in practice
\beq
\beta\gtrsim{1/\alpha}-1.
\eeq

\newpage
\clearpage

\clearpage

\clearpage

\clearpage

\begin{thebibliography}{99}
\bibitem{terms_sigmaV}
If only the first term on the right-hand side of Eq. \eqref{sigmaV} is retained, the Goldhaber-Gordon fit to the resulting $\sigma(V)$ yields the same $s$ value as the one for $\sigma(T)$; in this case, the effect of a low bias voltage $V$ is similar  to that of an effective temperature $T=\sqrt{3}/(2\pi) V$ [consistent with the relationship between $T$ and $V$ shown in Eq. \eqref{FL}] .
\bibitem{prefactor}
To evaluate the analytic $T_K$ we used a prefactor of $1.13$ in front of the right-hand side of Eq. \eqref{Kondo_scale_exact}, as indicated by NRG calculations (see Ref. \cite{Micklitz}); we also used minor corrections of order $1/u$ inside the exponential factor.
\end{thebibliography}
\end{document}